# Artificial Neural Network (ANN) - Oscillatory Neural Network (ONN) Hybrid System Using Domain-Wall Synapse Devices and Nano-Constriction Spin Hall Nano Oscillators

**Raman Hissariya**[1] **& Gajjala Venkata Sreekar Reddy**[1] **& Ashwin Tulapurkar**[1] **& Debanjan Bhowmik**[1,2]

[1]Department of Electrical Engineering, Indian Institute of Technology Bombay, Mumbai, Maharashtra 400076, India

[2]Centre for Semiconductor Technologies (SemiX), Indian Institute of Technology Bombay, Mumbai, Maharashtra 400076, India

E-mail:   debanjan@ee.iitb.ac.in

**Abstract.** A coupled spintronic oscillator array has been considered attractive for neuromorphic computing applications. Experimental reports have shown the nano-constriction geometry to be a relatively easier-to-fabricate platform for implementing such spin oscillators, but most prior reports on training and inference algorithms for neuromorphic computing using spin oscillators have been mostly restricted to the nano-pillar geometry. Also, those prior reports involve updating the natural frequency values of the oscillators and moving the synchronization regions on to the data clusters during the offline learning phase, which has associated challenges. In this context, we design and simulate a novel artificial neural network (ANN) - oscillator neural network (ONN) algorithm where in the offline learning phase, the weight parameters of the ANN are updated such that the data clusters are instead moved to the synchronization regions of spin Hall nano oscillators (SHNOs) in the nano-constriction geometry, as obtained through micromagnetic simulations. We further simulate the on-chip inference part of the ANN-ONN algorithm where the ANN is implemented on a crossbar array of domain-wall synapse devices, as simulated here through micromagnetics, and the ONN is implemented on nanoconstriction SHNOs. We show successful data classification for both binary and multi-class classification tasks to demonstrate the generalizability of our proposed scheme.

## 1. Introduction

Inspired by the fact that oscillatory activities have been recorded in the brain, oscillatory neural networks (ONN) have been proposed and implemented through emerging hardware to carry out machine learning and optimization tasks fast (preferably in real time) with high energy efficiency and with less hardware resources for edge artificial intelligence (AI) applications [1 - 9]. For example, as a major advantage of ONN, it has been shown that the number of weight parameters needed in an ONN is much less compared to the standard multi-layer perceptron (MLP) for the same data classification task [10]. This is because oscillators, acting as neurons, interact with each other, and their individual characteristics can also be modulated, unlike neurons in a standard MLP network [10].

A coupled spintronic oscillator array has been experimentally demonstrated to be a suitable neuromorphic platform to implement an ONN owing to the following properties spintronic oscillators exhibit: high frequency of operation (in GHz), low energy consumption, frequency tunability using DC current and DC magnetic field, ability to synchronize with each other and to RF magnetic fields and currents, etc [10 - 26]. However, initial experimental reports on spin-oscillator-based data classification [10, 13], both using the nano-pillar geometry and the nano-constriction geometry [27 - 29], have been limited to specific data sets without showing much generalizability for the classification schemes. Some simulation-based studies have though been reported with more generalizable data classification schemes (off-line learning on a conventional computer, inference on spin oscillator hardware using the nano-pillar geometry) have been reported by Vodenicarevic *et al* [30] and by Hemadri Bhotla *et al* [31]. In these reports, through off-line pre-processing done on a conventional computer, the data that is to be classified is first reduced from its original higher dimensional space to a lower dimensional space. Then again, by implementing an off-line learning/ training algorithm on the conventional computer, natural frequencies of the oscillators are updated and their final values are calculated such that the synchronization regions of the oscillators match with the data clusters (in the lower dimensional space) corresponding to the different data classes. Then, for inference/ testing, the natural frequencies of the actual oscillators are set as per the above calculated values (it's possible to apply different current densities for different spin oscillators in the nano-pillar geometry [27 - 29]). Provided that the synchronization regions from the physical spin oscillator hardware match with that predicted by offline computation at these calculated natural frequency values, inference can be carried out with high accuracy on the spin oscillator hardware using this scheme. However, a major challenge with the above strategy is that synchronization regions may not retain their original and desired shapes as the natural frequencies of the oscillators are adjusted because of the complicated nature of the coupling mechanisms among the oscillators. And if the shapes of the synchronization regions change, there is a major accuracy loss if the aforementioned offline learning algorithm is implemented. This is why data classification through this offline learning scheme for nano-pillar spin oscillators has been mostly limited to binary classification thus far. And a novel scheme is required to extend data classification to multiple classes using spin oscillators.

Also, as mentioned before, this offline learning scheme has been designed for spintronic oscillators in the nano-pillar geometry since tuning of natural frequencies through adjusting current density values is easier in this geometry [27 - 29]. However, fabrication of such oscillators is difficult given that fabrication of magnetic

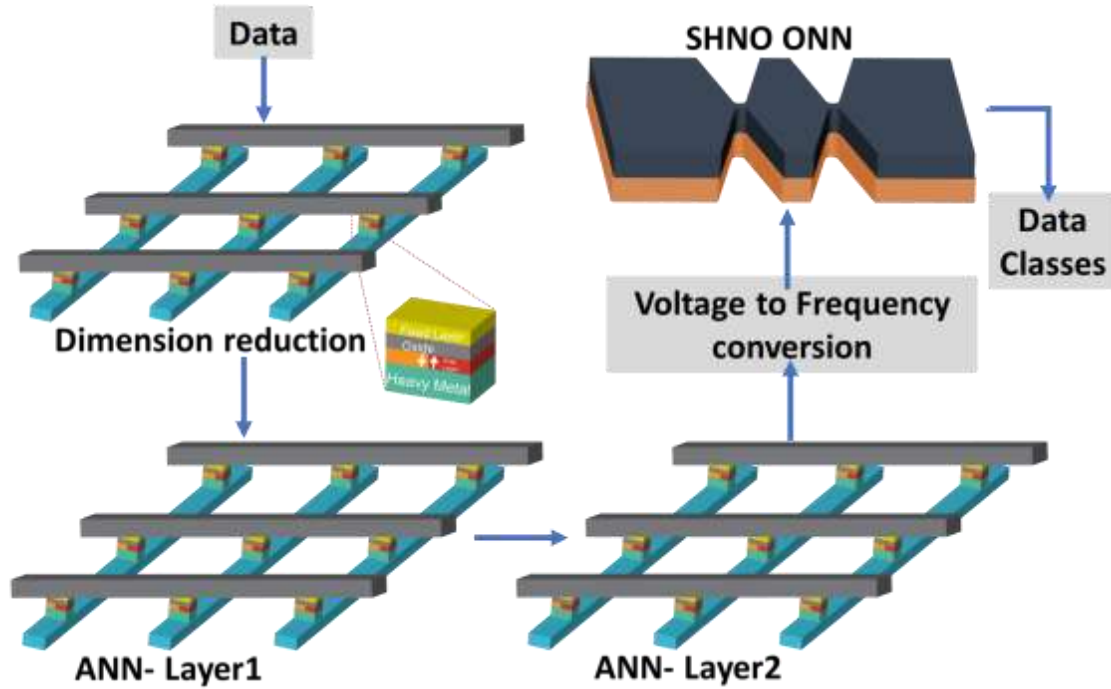

***Figure 1:*** *Schematic for carrying out inference using the proposed ANN-ONN hybrid algorithm. The required components are a crossbar array of domain-wall synapse devices to carry out dimension reduction of input data, two more such crossbar arrays for transformation of data into data clusters that align with synchronization regions of coupled SHNOs, and an array of coupled SHNOs in the nano-constriction geometry for the final classification*

tunnel junctions (MTJ) with a very thin insulating barrier is essential for their operation [29, 32, 33]. In comparison, spin Hall nano oscillators (SHNOs) in the nano-constriction geometry are much simpler to fabricate since they only need a heavy-metal-ferromagnetic-metal bilayer for the operation; in-plane current flowing through the device generates spin-polarized electrons at the interface of the heavy metal and the ferromagnetic metal and triggers oscillation in the ferromagnetic metal layer [12, 14, 15]. In recent reports, mutual synchronization has been experimentally reported for 64 SHNOs in one design and even 105,000 SHNOs in another design [14, 34, 15, 17, 21, 22, 23, 35]. But it's extremely difficult to vary the current densities through different nano-constrictions independently and modulate the natural frequencies of the nano-constriction SHNOs independently, which is required for the offline learning algorithm described above. This further requires the development of a novel data classification scheme for spin oscillators in the nano-constriction geometry.

In this article, we propose a modified oscillator-based data classification scheme so that it can be adapted to SHNOs in the nano-constriction geometry and it can be generalized to a wide range of data and classes. In the offline learning part of our proposed scheme, instead of moving the synchronization regions of the oscillators on to the data clusters (after dimension reduction) through adjustments of oscillators' natural frequencies (as described above), the ONN is complemented with an artificial neural network (ANN) so that dimension reduction of the original data can be carried out and the obtained data clusters automatically coincide with the synchronization regions of the nano-constriction SHNOs (obtained through micromagnetic simulations on mumax3 simulation package [36] in our work but can also be obtained experimentally [13]). As a result, natural frequencies of the oscillators do not need to be adjusted, which has major advantages for the nano-constriction geometry as discussed above.

For on-chip inference corresponding to this novel scheme, we have proposed using crossbar arrays of spintronic domain-wall synapse devices [37 - 47] to implement the ANN and the same nano-constriction SHNOs to implement the ONN, as shown in Fig. 1. Since both these devices involve similar materials stack (heavy-metal-ferromagnetic-metal-bilayer) and exploit similar physics (related to spin-orbit-torque), they are compatible with each other. To simulate on-chip inference, we have modelled both the nano-constriction SHNOs and domain-wall devices using micromagnetics (on mumax3 simulation package) and then incorporated these device-level results in our high-level language (Python) code for on-chip inference.

In this article, we report both binary classification and multi-class classification using this proposed scheme. For the purpose, we have explored the Fisher's Iris data set of flowers [48, 49] as well as the MNIST data set of handwritten digits [50, 51] and shown high classification accuracy numbers for both data sets. For Iris, we show both binary classification and three-class classification. For MNIST, utilizing three distinct synchronization regions of two nano-constriction SHNOs, we show three-class classification (images of digit 0 vs that of digit 1 vs that of digit 2).

In Section 2 of this article, we have discussed the auto-oscillation and synchronization properties of the nano-constriction SHNOs as we have obtained through micromagnetic simulations. In Section 3, we discuss our micromagnetic simulation results for the domain-wall synapse device to be used in the crossbar array for the above dimension reduction (data transformation). In Section 4, we discuss in details our proposed ANN-ONN hybrid system including its training process. We discuss data classification results (including accuracy numbers) on Iris and MNIST data sets based on simulation of the on-chip inference scheme of the designed ANN-ONN hybrid system on the array of SHNOs (designed in Section 2) and domain-wall synapse devices (designed in Section 3). In Section 5, we conclude the article.

## 2. Nano-constriction SHNOs: Modelling, Auto-Oscillation and Synchronization Characteristics

*2.1. Methodology: Modelling of Single SHNO*

In this article, we have modeled nano-constriction SHNO devices based on the platinum (Pt)/ permalloy (Py) bilayer with Pt thickness 6 nm and Py thickness 5 nm. Pt acts as the heavy metal, and Py acts as the ferromagnetic metal. The conductivity, relative permeability, and permittivity values for Pt have been considered to be $8.96 \times 10^6$ S/m, 1.000265, and 1 respectively. The conductivity, relative permeability, and permittivity values for Py have been considered to be $1.74 \times 10^6$ S/m, 8000, and 1 [15, 52, 53]. Using these parameter values, we have utilized COMSOL Multiphysics Software [54] to calculate the current density flowing through Pt and Py layers and the magnetic field generated due to Oersted effect in the nano-structure. Current density and Oersted magnetic field profiles obtained through COMSOL have been shown in Section S1 of Supplementary Material. Given that Pt has higher conductivity than Py, we observe that majority of in-plane current flows through the Pt layer. To enhance current density at a particular region of the device and trigger auto-oscillations in the ferromagnetic layer, a nano-constriction has been defined at the centre of the device with a 100 nm nanoconstriction width, an opening angle of 22 and a 50 nm radius of curvature (as shown in Fig. 2(a)).

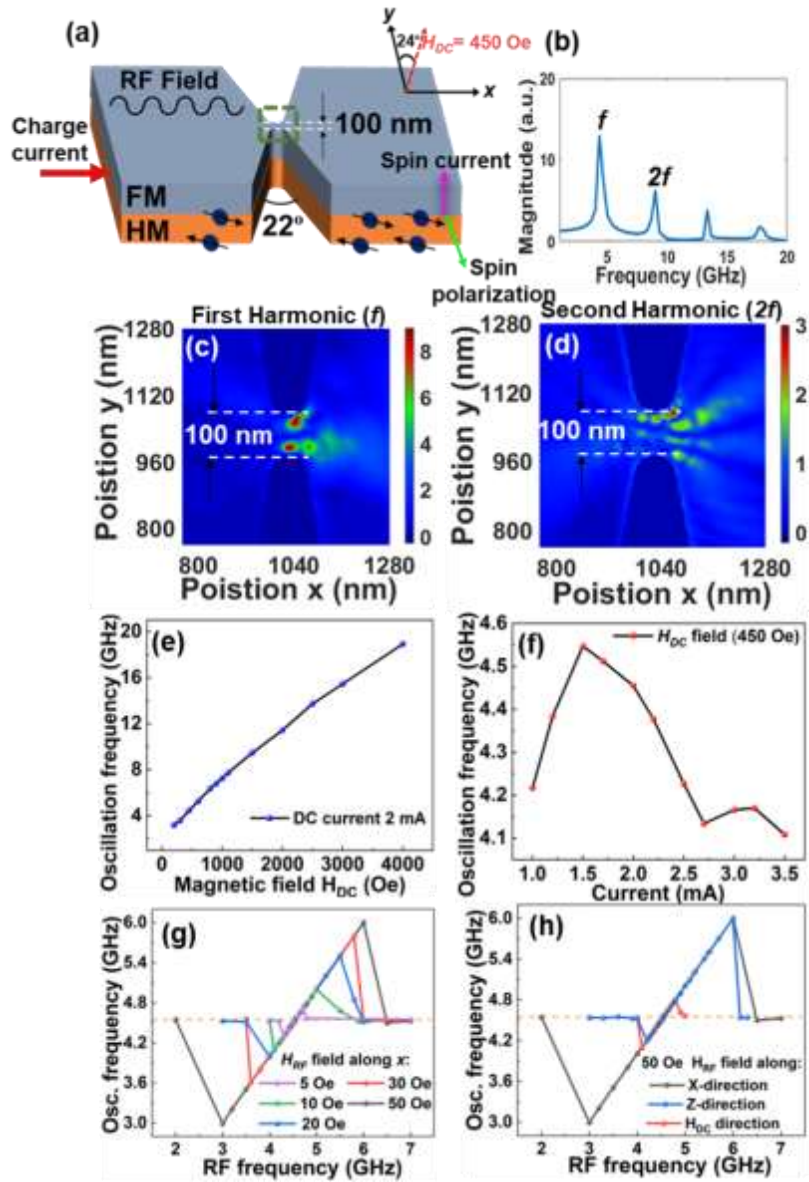

*Figure 2:* *(a) Device schematic of a single SHNO. (b) FFT of the variation of the average magnetic moment in z direction (out of plane) with time for the selected nano-constriction region in (a) (identified through dashed border line). (c) First harmonic mode of oscillation under 2.0 mA in-plane current: spatial distribution of FFT shown. (d) Second harmonic mode of oscillation under 2.0 mA in-plane current: spatial distribution of FFT shown. The color bars for both (c) and (d) correspond to power spectral density. (e) Variation of oscillation frequency of the SHNO with magnitude of DC field ($H_{DC}$). (f) Variation of oscillation frequency with DC current magnitude. (g) Variation of oscillation frequency with frequency of external RF field $H_{RF}$ (applied along the direction of DC current: x axis) for different RF field magnitudes. (h) Variation of oscillation frequency with frequency of RF field as the direction of RF field varies.*

***Table 1:*** *Materials parameter values used for micromagnetic simulation of the SHNOs*

| Parameter (Units) | Value |
|---|---|
| Saturation magnetization, $M_s$ (A/m) | $6 \times 10^5$ |
| Exchange strength, A (pJ/m) | 10 |
| gyromagnetic ratio, $\gamma$ (*rad/s.T*) | $1.855 \times 10^{11}$ |
| Gilbert damping, $\alpha$ | 0.02 |
| Spin Hall angle, $\theta_{SH}$ | 0.08 |
| Lambda, $\Lambda$ | 1 |

According to the physics of spin Hall effect, in these nanoconstriction SHNOs, in-plane charge current of current density $J_c$ generates spin current as follows: $J_s=\theta_{SHE}J_c$, under the assumption that the thickness of the heavy metal layer is higher than the spin diffusion length in the heavy metal, which is the case here since the thickness of the Pt layer is 6 nm and spin diffusion length inside Pt = 1-2 nm [55, 56]. Charge current distribution is obtained for our simulated nano-structure in COMSOL as mentioned above and then converted to spin current using the above expression.

The dynamics of the magnetic moments in the ferromagnetic layer (Py) under the influence of this spin current distribution and Oersted field distribution is next studied using the micromagnetics package mumax3 [36]. A 2048 nm × 2048 nm × 5 nm device, discretized into a grid of 512 × 512 × 1 cells (each cell with the dimension of 4 nm × 4 nm × 5 nm), has been studied on mumax3. A greyscale 2048 × 2048 pixel image mask has been used to carve out regions where the ferromagnet is absent and thereby implement the nano-constriction geometry (Fig. 2(a)).

Using the mumax3 package, we solve LLGS equation numerically for the magnetic moments of the ferromagnetic layer under the influence of the applied magnetic field and also Slonczewski spin torque arising out of the in-plane current flow, as given by:

$$\tau_{DL} m \times \sigma \times m, \tag{1}$$

where $\tau_{DL}$ and **m** are spin torque magnitude and reduced magnetization vectors respectively. Slonczewski spin torque $\tau_{DL}$ is considered to be [14]:

$$\tau_{DL} = \frac{\gamma j_e \hbar \epsilon}{e t \mu_0 M_S} \tag{2}$$

where $\gamma$, $M_s$, and $t$ are the gyromagnetic ratio, saturation magnetization, and ferromagnetic layer thickness values respectively. The spin polarization efficiency ($\epsilon$) in the expression for $\tau_{DL}$ is given by:

$$\epsilon = \frac{\theta_{SH} \Lambda^2}{(\Lambda^2+1)+(\Lambda^2-1)\vec{\sigma}.\vec{m}} \tag{3}$$

where the parameter $\Lambda$ determines the degree of angular dependency of the generated spin-orbit torque (generated as a result of the spin current) on the relative orientation between the magnetization direction and spin polarization direction [14]. This value has been set to 1 here to remove any angular dependency from the simulation. Other simulation parameter values, with respect to the above equations, have been taken from previous experimental reports [14] for the simulation work in this article and have been listed in Table 1.

Throughout the mumax simulations in this paper, the DC magnetic field has been applied at an angle of 66$^o$ with respect to the DC current direction (x axis) within the x-y plane (hence 24$^o$ with y axis), as shown in Fig. 2(a) [14]. Keeping this direction of the DC magnetic field unchanged, simulations have been performed while keeping the in-plane current magnitude constant and varying the DC field magnitude, as well as keeping the DC magnetic field magnitude constant while varying in-plane current value.

## 2.2. Results: Auto-Oscillation and Synchronization to RF Field in Single SHNO

We first study auto-oscillation characteristics of a single SHNO device, as shown in Fig. 2(a), in the presence of the aforementioned DC magnetic field but in the absence of any RF magnetic field. When 2 mA in-plane current flows through the device along x direction, the FFT of the time-dependent z-component of magnetization (out-of-plane component) averaged across the selected central constriction region (as identified through dashed boundary line in Fig. 2(a)) shows peaks at fundamental (4.4 GHz), second harmonic (8.9 GHz), and higher harmonics (Fig. 2(b)).

Next, in Fig. 2(c), (d), the amplitude corresponding to the above fundamental frequency and the above second harmonic frequency respectively is plotted as a function of x position and y position (axes shown in Fig. 2(a)). Bright red regions at the nano-constriction edges in these spatial FFT distributions correspond to peaks in power density, and show that magnetic oscillations mostly happen at these edges. Similar oscillations proximal to the edge have also been reported through simulations in the article by Kendziorczyk *et al* [57].

**Table 2:** *Variation of synchronization range of a single SHNO device with the amplitude and direction of RF magnetic field ($H_{RF}$)*

| Direction of RF field ($H_{RF}$) | Amplitude of RF field ($H_{RF}$) | Sync. range (GHz) |
|---|---|---|
| x | 5 Oe | 0.4 |
| x | 10 Oe | 1.0 |
| x | 20 Oe | 1.5 |
| x | 30 Oe | 2.2 |
| x | 50 Oe | 3.0 |
| z | 50 Oe | 1.8 |
| same as $H_{DC}$ (in xy plane, 66° with x, 24° with y) | 50 Oe | 0.7 |

Next, the DC magnetic field is varied in magnitude keeping its direction the same as mentioned in the previous sub-section; the DC current magnitude is also kept fixed at 2 mA. FFT peak for the first harmonic (natural frequency) of the SHNO's oscillation is plotted as a function of DC magnetic field magnitude in Fig. 2(e) and is found to increase monotonically with DC field magnitude. However, when the DC field is fixed at 450 Oe (same direction as before) but the DC current magnitude is varied instead, natural frequency doesn't show any monotonic dependence on current magnitude (Fig. 2 (f)). This is not unexpected and has also been observed in reports by [15, 58, 57, 17]. Plots of power spectral density as a function of frequency and magnitude of externally applied DC magnetic field, corresponding to Fig. 2(e) and (f), are shown in Section S2 of Supplementary Material.

Next, while applying DC field of 450 Oe along the same direction as before and DC current of 2 mA along the same direction as before (x axis), external RF magnetic field is also applied. In Fig. 2(g), keeping the RF field direction fixed along x, RF magnetic field strength is varied. In Fig. 2(h), keeping the RF field magnitude fixed at 50 Oe, RF magnetic field's direction is varied.

Based on the SHNO's first harmonic frequency vs RF field's frequency plots in Fig. 2(g), (h), we identify the synchronization range of the SHNO. Within the synchronization range, frequency of first harmonic of the SHNO is equal to the RF field's frequency (which satisfies the definition of synchronization). Outside the synchronization range, frequency of first harmonic of the oscillator follows the natural frequency of the SHNO, which is the same as the first harmonic frequency in the absence of any RF field and has already been plotted in Fig. 2(e), (f) and discussed before. Synchronization ranges for different RF field magnitudes and directions, as obtained from Fig. 2(g), (h), are listed in Table 2. Synchronization range increases with RF field magnitude as expected. Also, the synchronization range of the SHNO is found to be the maximum when RF field is applied along the x direction (Fig. 2(h)). We will utilize this finding while synchronizing multiple adjacent SHNOs next.

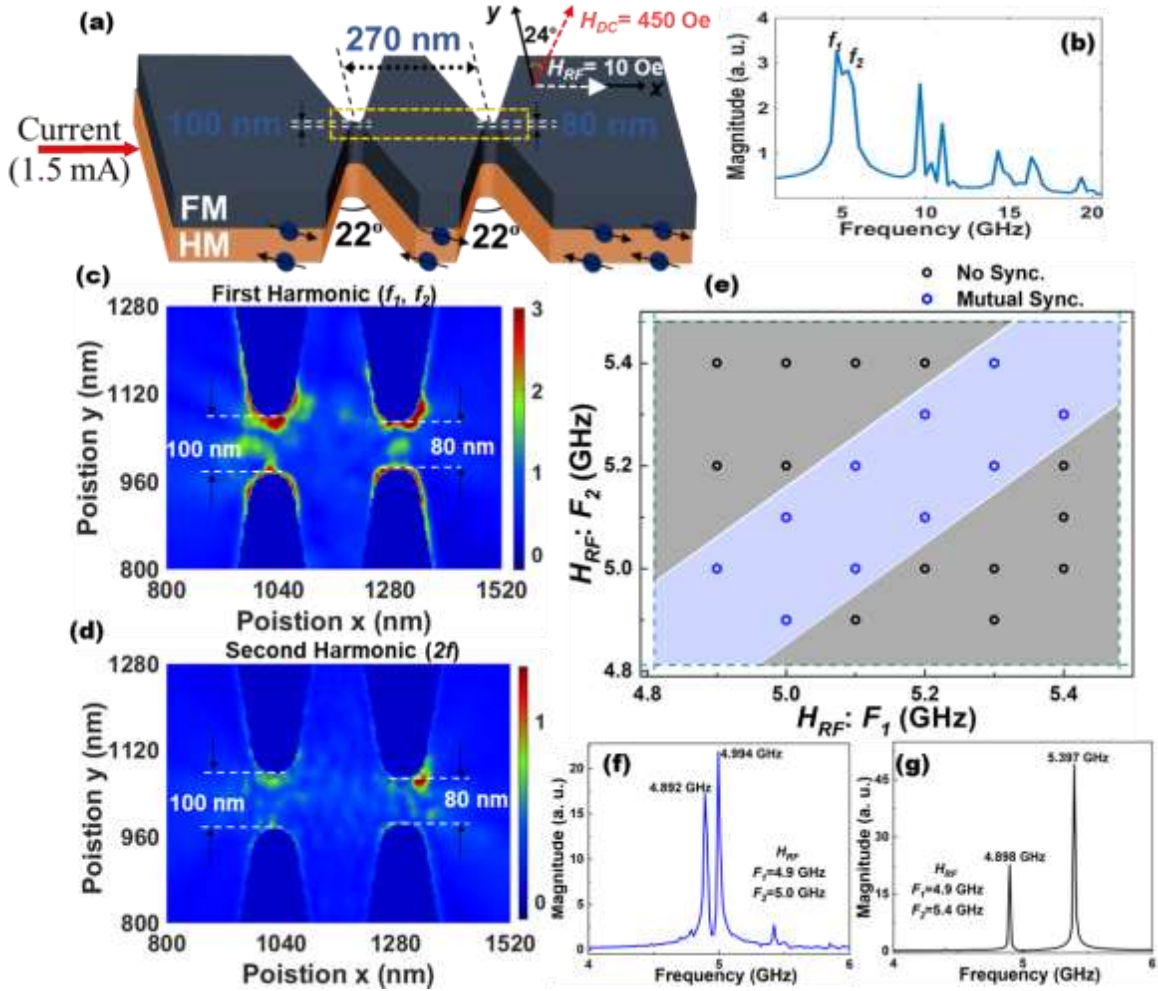

***Figure 3:*** *(a) Device schematic of two coupled SHNOs with RF fields applied along x axis. (b) With 1.5 mA DC current applied, FFT of time-dependent out-of-plane magnetization component ($m_z$) in the selected nano-constriction region in (a) (identified through dashed yellow border line). (c), (d) Power spetral density from FFT at first and second harmonic frequencies plotted vs position coordinates x and y (e) Synchronization map, intended for binary classification, obtained by applying RF magnetic fields of frequencies $F_1$ and $F_2$ along the x axis. (f) FFT spectrum when $F_1$ = 4.9 GHz, $F_2$ = 5 GHz. Since the fundamental frequencies of the two SHNOs (4.892 GHz, 4.994 GHz) differ by less than 120 MHz, the two SHNOs are considered synchronized (blue region in (e)) as per our definition. (g) FFT spectrum when $F_1$ 4.9 GHz, $F_2$ 5 .4 GHz. Since the fundamental frequencies of the two SHNOs now differ by much more than 120 MHz, the two SHNOs are not considered synchronized (grey region in (e)).*

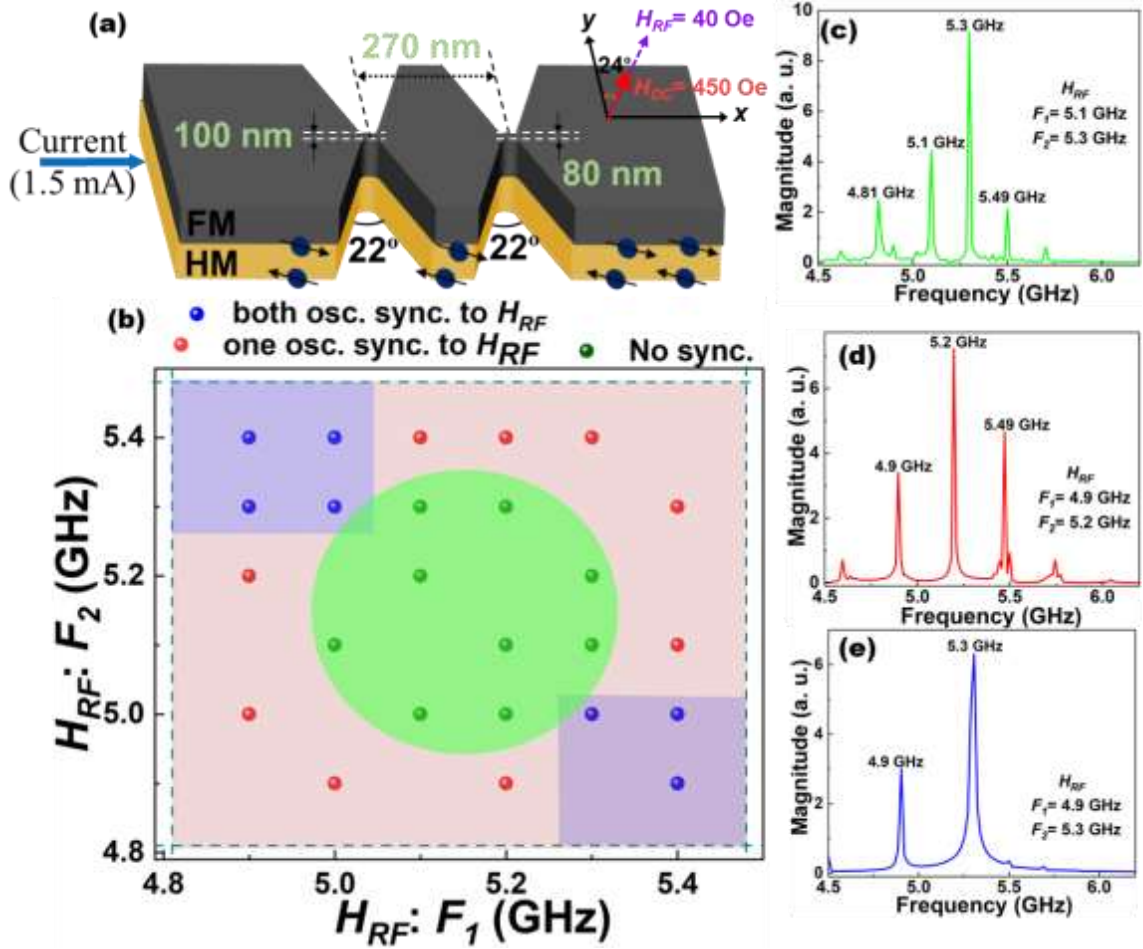

***Figure 4:*** *(a) Device schematic of two coupled SHNOs with RF fields applied in the direction of DC magnetic field (in xy plane, 66° with x axis, 24° with y axis) (b) Synchronization map, intended for three-class classification, obtained by applying RF magnetic fields of frequencies $F_1$ and $F_2$ along the direction of DC field. (c) FFT spectrum from the nano-constriction region when $F_1$ = 5.1 GHz and $F_2$ = 5.3 GHz shows four peaks. This corresponds to light green region in (b). (d) FFT spectrum when $F_1$ = 4.9 GHz and $F_2$ = 5.2 GHz shows three peaks. This corresponds to light red region in (b). (e) FFT spectrum when $F_1$ = 4.9 GHz and $F_2$ = 5.3 GHz shows two peaks. This corresponds to light blue region in (b).*

*2.3. Methodology: Modeling of Two Coupled SHNOs*

Next, we have carried out micromagnetic simulation of two adjacent SHNOs in the nano-constriction geometry, as shown in Fig. 3 (a) (same schematic repeated in Fig. 4 (a), only difference lies in the direction of RF field). The constriction widths of the two nano-constrictions are chosen to be slightly different so that their natural frequencies are slightly different and they can be forced into and out of synchronization by varying the applied RF magnetic fields. The distance between the two constrictions is chosen to be 270 nm. Uniform in-plane DC magnetic field of 450 Oe is applied along the same direction as that applied earlier for the single SHNO device (confined to x-y plane, making 24° angle with y axis).

For the device schematics of Fig. 3 (a) (same as Fig. 4 (a)in the absence of RF field), if DC current of 1.5 mA is applied along the x axis, auto-oscillations are induced in the two nano-constrictions at slightly different frequencies as mentioned above and as evident from FFT spectrum averaged the two nano-constriction regions ($f_1$, $f_2$ in Fig. 3 (b)). Spatial distribution of FFT intensity values in this frequency range shows oscillations again happening at the edges of the nano-constrictions like before (Fig. 3 (c), (d)).

Further, two external RF magnetic field signals of different frequencies $F_1$ and $F_2$ are both applied along the x axis (Fig. 3 (a)) to generate the synchronization map of Fig. 3 (e), intended for binary classification. In another case, two external RF field signals of different frequencies $F_1$ and $F_2$ are applied both along the direction of the DC magnetic field (confined to x-y plane, making 66° angle with x axis and hence 24° angle with y axis: Fig. 4 (a)) to generate the synchronization map of Fig. 4 (b), intended for multi-class classification. We discuss the features of these two syncyhronization maps next.

## 2.4. Results: Binary Classification with Two Coupled SHNOs

Two RF fields, each with an amplitude of 10 Oe but of different frequencies $F_1$ and $F_2$, are applied along the x axis in Fig. 3 (a) for the purpose of binary classification. $F_1$ and $F_2$ are varied. For different combinations of $F_1$ and $F_2$, FFT averaged across the two nano-constriction regions is computed (after letting the system evolve for 100 ns in each case) and the two fundamental/ first harmonic frequencies of the two regions ($f_1$, $f_2$) are obtained. If $|f_1 - f_2| \leq 120$ MHz, we consider the two SHNOs synchronized. Such a case is shown in Fig. 3 (f); in this case, applied frequencies $F_1$ = 4.9 GHz, $F_2$ = 5 GHz. If $|f_1 - f_2| > 120$ MHz, we consider the SHNOs not synchronized, as shown in Fig. 3 (g) for example (here, $F_1$ = 4.9 GHz, $F_2$ = 5.4 GHz).

The cases, or $F_1$, $F_2$ combinations, which lead to synchronization by the above definition, are marked with blue circles in the synchronization map of Fig. 3 (e) ($F_1$ and $F_2$ are its two axes). The cases which lead to lack of synchronization are marked with black circles in Fig. 3 (f). In this process, the synchyronization map for binary classification is obtained with two SHNOs synchronized corresponding to one category of data, as shown in the later sections (blue circles in light blue background) and with two SHNOs not synchronized corresponding to another category of data (black circles in grey background).

## 2.5. Results: Three-Class Classification with Two Coupled SHNOs

For three-class classification, two RF fields, each with an amplitude of 40 Oe but of different frequencies $F_1$ and $F_2$, are applied along the direction of DC field (confined to x-y plane, making 24° angle with y axis) in Fig. 4 (a). $F_1$ and $F_2$ are varied. For different combinations of $F_1$ and $F_2$, FFT averaged across the two nano-constriction regions is computed (after letting the system evolve for 100 ns in each case) and the peaks in the FFT spectrum are obtained. Three different cases are henceforth identified:

(i) For some combinations of ($F_1$, $F_2$), four distinct peaks, corresponding to the two frequencies of external RF field ($F_1$) and ($F_2$) and first harmonic frequencies of the two SHNOs ($f_1$ and $f_2$), are observed in the FFT spectrum. Such a case is shown in Fig. 4 (c): green plot. This means neither SHNO is synchronized to any RF field nor are the SHNOs synchronized with each other. This region is identified as light green region with green circles in the three-class synchronization map of Fig. 4 (b).

(ii) For some other combinations of ($F_1$, $F_2$), three distinct peaks are observed in the FFT spectrum: Fig. 4 (d) (red plot). This means one SHNO is synchronized to one RF field (one common peak instead of two) and the other SHNO is not synchronized to any RF field or to the first oscillator (two other peaks hence, one from the other SHNO, another from the other RF field). This region is identified as light red region with red circles in the three-class synchronization map of Fig. 4 (b).

(iii) For some other combinations of ($F_1$, $F_2$), only two distinct peaks are observed: Fig. 4 (e) (blue plot). Here, one SHNO is synchronized to one RF field (one common peak), and another SHNO is synchronized to another RF field (another common peak). This region is identified as blue region with blue circles in the three-class synchronization map of Fig. 4 (b).

| Parameter (Units) | Value |
|---|---|
| Saturation magnetization, $M_s$ (A/m) | $9.5 \times 10^5$ |
| Exchange strength, $A$ (pJ/m) | 15 |
| Anisotropy constant, $K_{u1}$ | $10^6$ |
| Gilbert damping, $\alpha$ | 0.3 |
| Spin Hall angle, $\theta_{SH}$ | 0.08 |

***Table 3:*** *Materials parameters used for micromagnetic simulation of the domain-wall synapse device*

.
## 3. Domain-Wall Device: Modelling and Synaptic Characteristic

### 3.1. Modelling Methodology

The domain-wall device is used in this article as a non-volatile memory (NVM) synapse device in the crossbar arrays of Fig. 1 to implement ANN. The physics behind operation of such a device has been described in detail in previous simulation-based and experimental reports [37 - 40, 42, 46, 47, 59 - 66]. We repeat the working principle briefly here. In this device, as shown in Fig. 5(a), an in-plane current generates spin-orbit torque (SOT) at the heavy-metal-ferromagnetic-metal interface, driving a domain wall in the free layer of the associated magnetic tunnel junction (MTJ) structure. As the domain wall moves, the MTJ conductance changes due to tunneling magneto-resistance (TMR), and thus, multiple electrically controlled synaptic states, corresponding to different domain-wall positions, can be modulated in the device. In Fig. 5(a), the purpose of the antiferromagnet at the edge is to pin the domain wall if it comes to the edge and prevent it from getting destroyed.

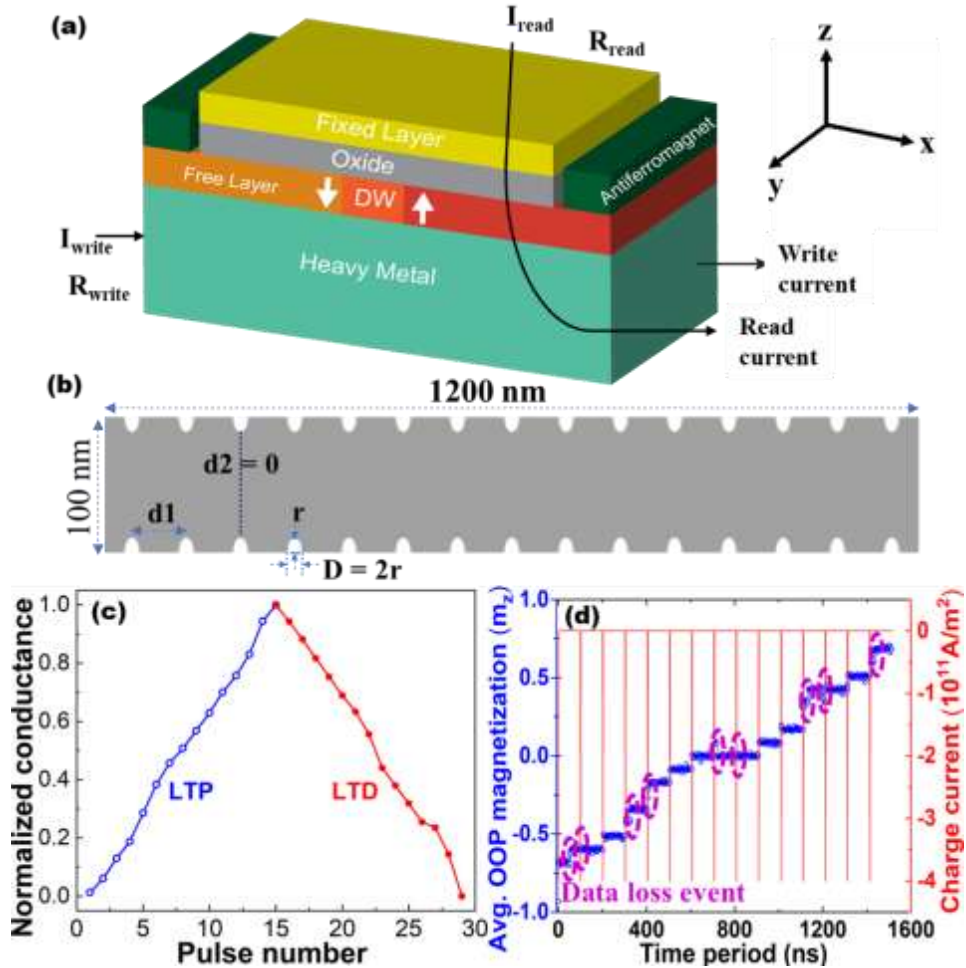

*Figure 5:* *(a) Schematic of the domain-wall synapse device. (b) Ferromagnetic free layer of the device (in which the domain wall moves) with circular notches (radius: 5.65 nm) on the edges, 60 nm apart from each other. (c) LTP and LTD characteristics of the device. (d) Retentivity of each synaptic state with 100 ns waiting time between two pulses. Data loss events are identified through circles in the plot.*

Recent experimental and simulation-based reports show that the addition of geometrical structures like meanders and notches at the edges of the device stabilizes the synaptic states through domain-wall pinning and improves the retentivity of the states as well as improves the linearity and symmetry of conductance control (long-term potentiation (LTP), long-term depression (LTD)) in the device [39, 62, 67, 68, 69].

In this article, we micromagnetically simulate a domain-wall synapse device using the same micromagnetic package mumax3 as before [36]. Circular notches are incorporated along both longitudinal edges of the device (1200 nm × 100 nm ferromagnetic layer), as shown in Fig. 5(b). Inside the notches, saturation magnetization $M_s$ is taken to be 0 to signify that the magnetic material is etched out in the notches. These circular notches each have a radius of 5.65 nm and are spaced 60 nm apart. Such positioning has been shown to be optimum in previous simulation-based studies [68]. Relevant simulation parameters for the ferromagnetic layer are listed in Table 3.

*3.2. Results: Synaptic Characteristics*

In our simulations, current pulses of current density higher than $4 \times 10^{11}$ A/m$^2$ and of duration 1 ns are applied to obtain the synaptic characteristics of the device. Current below this magnitude does not depin the domain wall from the notch and doesn't lead to domain-wall motion. A DC in-plane magnetic field of 3000 G is required to promote deterministic domain-wall motion, which is the signature of SOT. In our simulation, positive current pulses (current flowing in +x direction) move the domain wall along +x with respect to the device schematic of Fig. 5(a). This changes normalized magnetization (along *z* axis) from +1 to -1 and conductance of the vertical MTJ structure from minimum value to maximum value (normalized conductance from 0 to 1 in Fig. 5(c)). This results in the long-term potentiation (LTP) characteristic of the device. For long-term depression (LTD), negative current pulses are applied (current flowing in the -x direction), which move the domain wall in the reverse direction. As a result, the normalized magnetization (along z) gradually switches from -1 to +1, and the normalized MTJ conductance decreases from 1 to 0, resulting in LTD characteristic (Fig. 5(c)).

The LTP and LTD characteristics of the device are found to be sufficiently linear (conductance varies linearly with pulse number: Fig. 5(c)) owing to the presence of the notches. The LTP and LTD characteristics are also symmetric with respect to each other for the same reason. This reduces the accumulated error in the proposed ANN and improves the classification accuracy of the proposed ONN-ANN hybrid system. In Fig. 5(d), negative current pulses of density $4 \times 10^{11}$ A/m$^2$ and duration 1 ns are applied with waiting time 100 ns between the pulses. Fig. 5(d) shows that the domain wall rarely moves when current pulses are not applied because notches pin the domain wall and prevent its motion due to stray field, thermal noise, etc., which are included in the simulation. This significantly reduces data loss events and improves the retentivity of the synaptic states, as required in our proposed algorithm. Given that we have 15 notches at each longitudinal edge of the device (Fig. 5(a)) and thereby obtain 16 stable synaptic states as a result of 16 identical write pulses (Fig. 5(b), (c)), we conclude that our domain-wall synapse device is of 4-bit weight resolution.

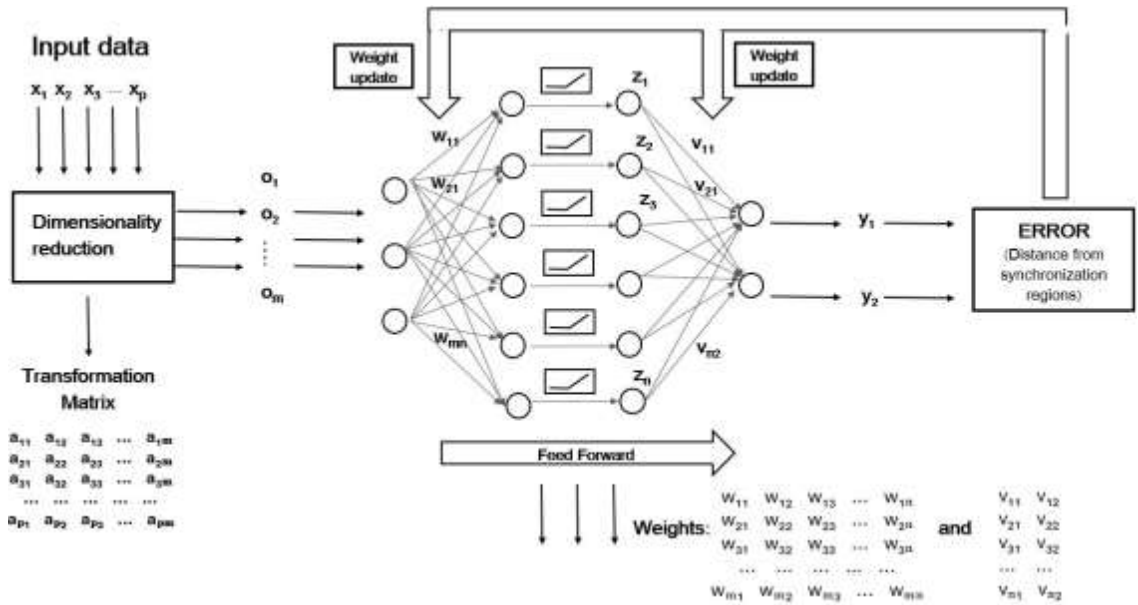

***Figure 6:*** *Detailed design of the proposed ANN that transforms the input data to two scalar values $y_1$ and $y_2$, which will be applied as frequency values of RF fields on the SHNO-based ONN during on-chip inference*

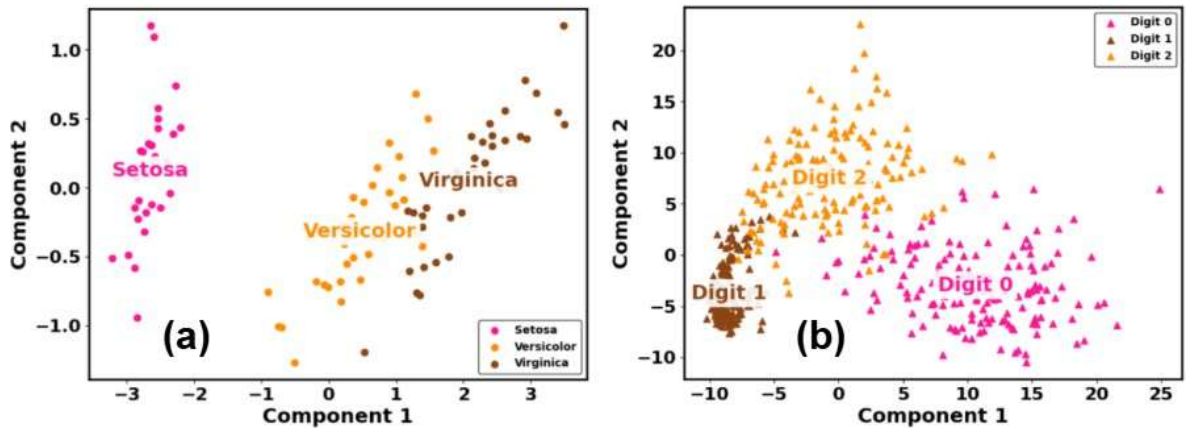

***Figure 7:*** *Projection of data from (a) the Iris data set and (b) the MNIST data set into a two-dimensional space using principal component analysis (PCA). Both train and test samples are used for this purpose.*

## 4. ANN-ONN Hybrid System: Training and Inference

As mentioned in Section 1, the purpose of our proposed ANN-ONN hybrid system is to transform the input data into reduced dimensions such that the data clusters map on to the different synchronization regions of the SHNOs (Fig. 3(e): 2 output classes, Fig. 4(b): 3 classes). We present the mathematical details of different stages of training and inference/ testing of this ANN-ONN system below. For Iris, 120 samples are used for training: 40 for Setosa category of flower, 40 for Versicolor, 40 for Virginica. 30 samples are used for testing, 10 for each category. For MNIST, 480 samples are used for training: 160 for digit 0, 160 for digit 1, and 160 for digit 2. 120 samples are used for testing, 40 for each of the three digits.

*4.1. Design of ANN for data transformation and data sets used*

Let the input vector corresponding to each data sample be represented as

$$\mathbf{x} = [x_1, x_2, \ldots, x_p] \in R^p, \tag{4}$$

where $p$ denotes the dimensionality of the original dataset. In our case, for Iris data set, $p = 4$ (4 features for each flower: petal length, petal width, sepal length, sepal width). For MNIST data set, $p = 784$ (28 x 28 pixels in each image).

At the first stage of our proposed ANN, we reduce this data vector to a lower-dimensional representation of dimension $m < p$, as shown in Fig. 6. For this purpose, principal component analysis (PCA) [70] is employed to compute a linear transformation matrix

$$\mathbf{A} \in R^{(p \times m)}, \tag{5}$$

such that the reduced-dimensional representation is given by

$$\mathbf{o} = [o_1, o_2, \ldots, o_m] = \mathbf{xA}. \tag{6}$$

In this work, $m = 2$ both for Iris data set and MNIST data set. Only train samples are used to come up with the matrix **A** for Iris or MNIST. Subsequently, the values of all elements of **A** are kept unchanged during the rest of the training process and also during inference process for all data samples.

Each PCA-reduced data sample is subsequently mapped to appropriate frequency values for the SHNO synchronization maps using a feedforward ANN. For Iris, the ANN consists of an input layer with 2 neurons, a hidden layer of 128 neurons, and an output layer with 2 neurons (since there are 2 frequency values in the synchronization maps). A rectified linear unit (ReLU) activation function is employed in the hidden layer to introduce nonlinearity in the mapping [70]. The hidden layer computes:

$$Z_j = ReLU\left(\sum_{i=1}^{m} w_{ij} O_i\right),$$
$$j=1, 2, \ldots n$$
$$(7)$$

where $n = 128$ and $\mathbf{W} \in R^{m \times n}$ is the hidden-layer weight matrix ($w_{ij}$ represents element of this matrix $\mathbf{W}$). The ReLU activation is defined as

$$\text{ReLU}(x) = \max(0, x). \tag{8}$$

The output layer computes:

$$Y_k = \sum_{j=1}^{n} v_{jk} Z_j, \quad k=1, 2 \tag{9}$$

where $\mathbf{V} \in R^{n \times 2}$ is the output-layer weight matrix ($v_{jk}$ represents each element of this matrix $\mathbf{V}$). The outputs $y_1$ and $y_2$ are then scaled to the frequency values $y_{F1}$ and $y_{F2}$ with respect to the SHNO synchronization maps. For MNIST, the same procedure is followed for the ANN designed as for Iris the only difference being there are two hidden layers in our ANN for MNIST: first one of 8 neurons, second one of 16 neurons.

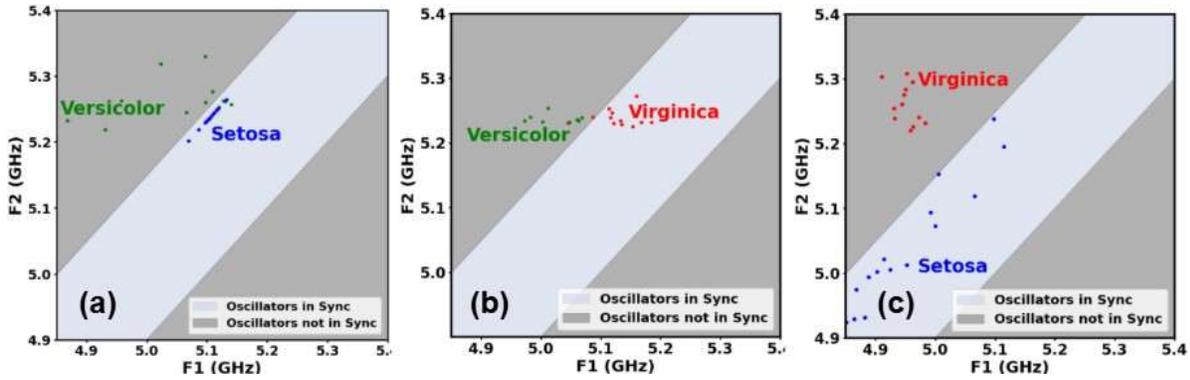

***Figure 8:*** *Binary classification on Iris test samples (inference) based on simulation of the domain-wall-synapse-SHNO ANN-ONN hybrid system: (a) Setosa vs Versicolor: Test accuracy = 95%; (b) Virginica vs Versicolor: Test accuracy = 90%; (c) Setosa vs Virginica: Test accuracy = 96.67%.*

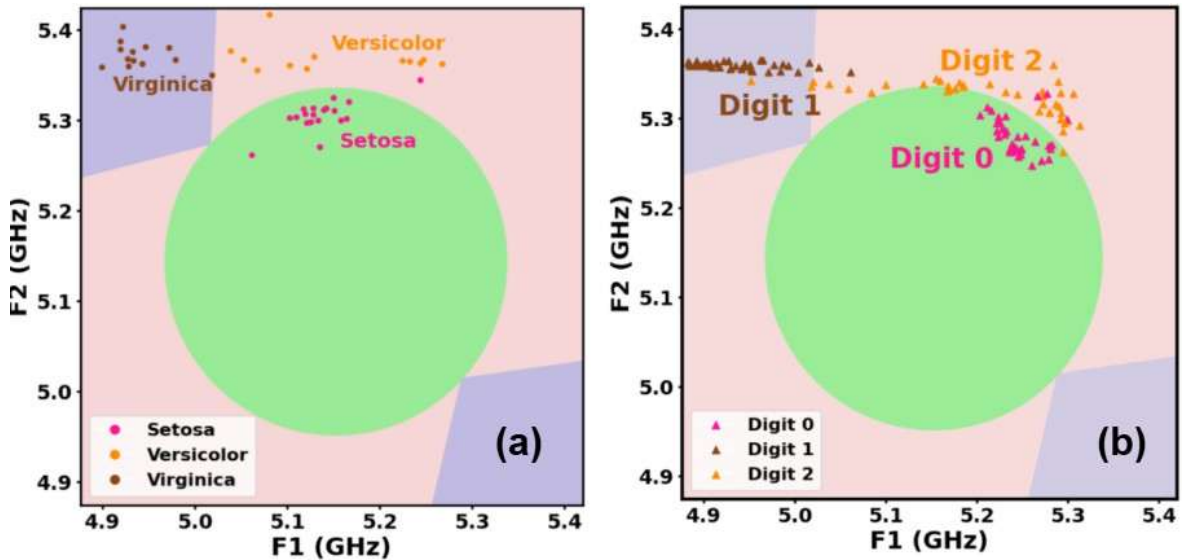

***Figure 9:*** *Multi-class classification on test samples (inference) based on simulation of the domain-wall-synapse-SHNO ANN-ONN hybrid system:(a) Iris: Test accuracy = 91.33% (b) MNIST: Test accuracy = 90.83%*

*4.2. Mapping SHNO Synchronization Regions to Output Classes*

To calculate the loss function for the above ANN (this loss needs to be minimized during training), first mapping of different SHNO synchronization regions to the different output classes needs to be carried out. For binary classification on the Iris data set (Fig. 8), mapping on the SHNO synchronization map of Fig. 3(e) is carried out as follows: the region where two SHNOs synchronize (light blue region) is mapped to one output class and the region where they don't (grey region) is mapped to another output class. For three-class classification on Iris (Fig. 9(a)) and MNIST (Fig. 9(b)), the region where both SHNOs synchronize to RF fields (light blue region) maps to Virginica category (for Iris) and Digit 1 (for Roshambo), the region where only one SHNO synchronizes to RF field (light red region) maps to Versicolor (for Iris) and Digit 2 (for Roshambo), and the region where no SHNOs synchronize to no RF field (light green region) maps to Setosa (for Iris) and Digit 0 (for Roshambo).

*4.3. ANN: Training*

The above ANN is trained through weight parameter update by minimizing the mean-squared-error loss function of the ANN ($E$) to which the error for each training sample contributes as follows:

$$\tfrac{1}{2}\sum_{k=1}^{2} (y_{Fk} - \hat{y}_{Fk})^2,$$

where $\hat{y}_{Fk}$ corresponds to the centroid of the SHNO synchronization region associated with the output class for the given data sample as identified above. This choice of loss function enables samples from each class to map toward the center of the corresponding synchronization region, thereby improving robustness to small perturbations. The weight parameter values of the ANN are updated by applying gradient descent on the mean-squared-error loss function with respect to the train samples [70]. The weight updates are hence given by:

$$\Delta w_{ij} = -\eta \frac{\partial E}{\partial w_{ij}}, \qquad \Delta v_{jk} = -\eta \frac{\partial E}{\partial v_{jk}} \qquad (11)$$

where $\eta$ is the learning rate (we choose $\eta$ = 0.05 for Iris, $\eta$ = 0.01 for MNIST). The updated weights are:

$$w_{ij}^{new} = w_{ij} + \Delta w_{ij}, \qquad v_{jk}^{new} = v_{jk} + \Delta v_{jk}. \qquad (12)$$

All network parameters and their values, as discussed above, are summarized in Table 4. We have implemented the above ANN training method on a conventional computer (CPU) through programming on a high-level language for this work. Train accuracy numbers obtained thereby for the aforementioned classification tasks are listed in Table 5. The number of training samples used is also mentioned. All training accuracy numbers are very high thereby showing successful training of the ANN.

*4.4. ANN-ONN: Inference on Domain-Wall-Synapse-SHNO Hardware*

Once the ANN is trained as above, we propose to implement the ANN with final weight parameters on the crossbar array of domain-wall synape devices and the ONN on the nanoconstriction SHNO hardware, as shown in Fig. 1, for inference/ testing. In this work, we simulate this inference process in our Python code by incorporating the LTP and LTD characteristics of the simulated domain-wall synapse device of Fig. 5(c) in the weight parameter values of the ANN. Thus, non-idealities of the hardware are now incorporated in our code. Given that each domain-wall synapse device has 16 conductance levels (4 bits of weight resolution) (Fig. 5(c)), we use a pair of such devices as a synapse unit in order to have synaptic weights of 8 bit resolution. In terms of obtaining a high accuracy number in this proposed inference scheme, synaptic weights of 4 bits are insufficient. So we make this design choice.

For the different aforementione classification tasks, test samples, after transformation through the simulated ANN incorporating the synapse device properties, are plotted with respect to the SHNO synchronization maps (obtained earlier) in Fig. 8 and Fig. 9. Inference/ test accuracy numbers obtained thereby on the test samples for the same classification tasks are listed in Table 5. More the number of test samples of one category lie in the synchronization region of another category, higher is the drop in test accuracy.

Our test accuracy numbers are slightly lower than our train accuracy numbers, which is common in machine learning. Difference between train and test accuracy is higher in the case of Iris compared to MNIST because there are only 30 test samples in Iris. So, wrong classification of even one test sample leads to significant accuracy drop for Iris.

| Network Parameters | Values |
|---|---|
| Structure of ANN for Iris | 2 × 128 × 2 |
| Learning ratio ($\eta$) for Iris | 0.05 |
| Structure of ANN for MNIST | 2 × 8 × 16 × 2 |
| Learning ratio ($\eta$) for MNIST | 0.01 |

*Table 4: Parameters for the designed ANN for Iris and MNIST data sets*

| Classification Task | Number of Train Samples | Number of Test Samples | Train Accuracy | Test Accuracy |
|---|---|---|---|---|
| *Binary Classification: Iris* | | | | |
| Setosa vs. Versicolor | 80 | 20 | 98.00% | 95.00% |
| Virginica vs. Versicolor | 80 | 20 | 100.00% | 90.00% |
| Setosa vs. Virginica | 80 | 20 | 100.00% | 96.67% |
| *Three-Class Classification* | | | | |
| Iris | 120 | 30 | 96.00% | 91.33% |
| MNIST (Digit 0 va 1 va 2) | 480 | 120 | 91.67% | 90.83% |

*Table 5: Classification performance of the proposed ONN-ANN hybrid system (both training set and test set are equally balanced among the used categories)*

## 5. Conclusion

Thus, in this article, we have designed and simulated an ANN-ONN hybrid system using spintronic domain-wall synapse devices and nanoconstriction-based SHNOs to carry out binary data classification on Iris data set and multi-class classification on Iris and MNIST. For this purpose, we have carried out a combination of micromagnetic simulation (mumax3 package) of domain-wall devices and SHNOs and system-level simulation of the ANN-ONN using high-level programming in Python, while incorporating our mumax3-based results in them. Our proposed method makes data classification through spintronic oscillators generalizable to a wide range of data classification problems and paves the way for efficient neuromorphic computing through spintronic oscillators.

**Supplementary Material**

Supplementary Material accompanying this article contains plots of current density profile, Oersted field profile, and power spectral density of the single nano constriction SHNO device, as obtained through COMSOL simulations and micromagnetic simulations.

## Acknowledgments

Raman Hissariya acknowledges support from Indian Institute of Technology Bombay through Institute Post Doctoral Fellowship (IPDF). Ashwin Tulapurkar and Debanjan Bhowmik acknowledge financial support from Ministry of Education India (MoE) through the MoE-STARS program (MoE-STARS/STARS-2/2023-0257). Debanjan Bhowmik also acknowledges support from Anusandhan National Research Foundation (ANRF) through ANRF-MATRICS program (MTR/2023/000851).

**Supplementary Material for "Artificial Neural Network (ANN) - Oscillatory Neural Network (ONN) Hybrid System Using Domain-Wall Synapse Devices and Nano-Constriction Spin Hall Nano Oscillators"**

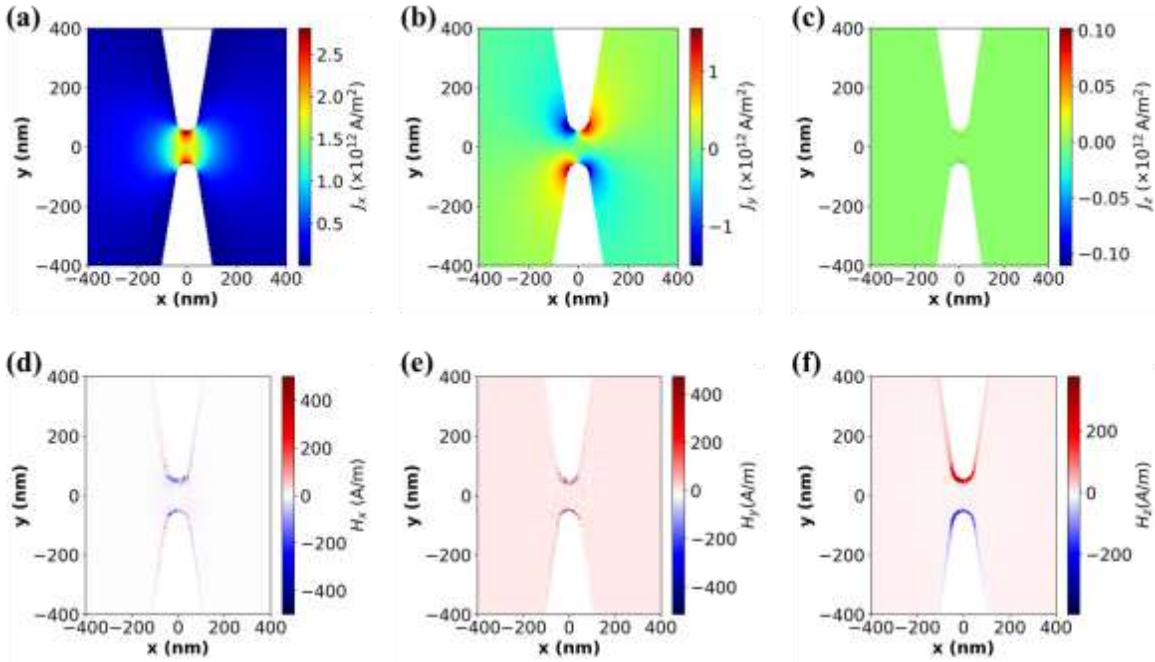

***Figure S1:*** *(a), (b), (c) Current density profile inside the heavy metal layer in x, y, z directions respectively. (d), (e), (f) In-plane magnetic field profile inside the ferromagnetic metal layer in x, y, and z directions.*

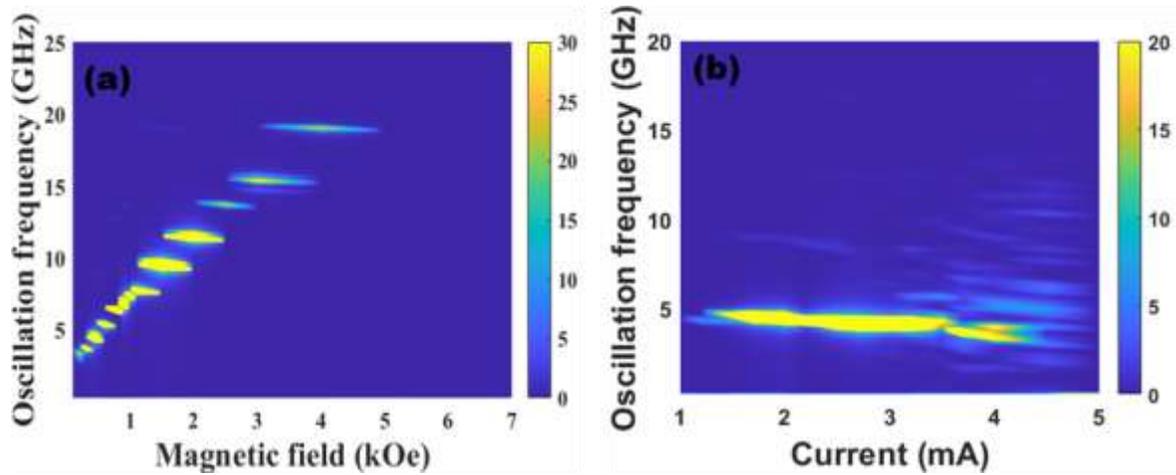

***Figure S2:*** *(a) Power spectral density (PSD) of the time-dependent z-component of the magnetization (out-of-plane magnetization component) as a function of applied DC magnetic field and frequency when a fixed DC current of 2 mA flows in the x direction (with respect to single nanoconstriction SHNO schematic of Fig. 2 (a) of the main text). (b) PSD of the time-dependent z-component of the magnetization as a function of applied DC current and frequency when DC field of 450 Oe is applied throughout.*

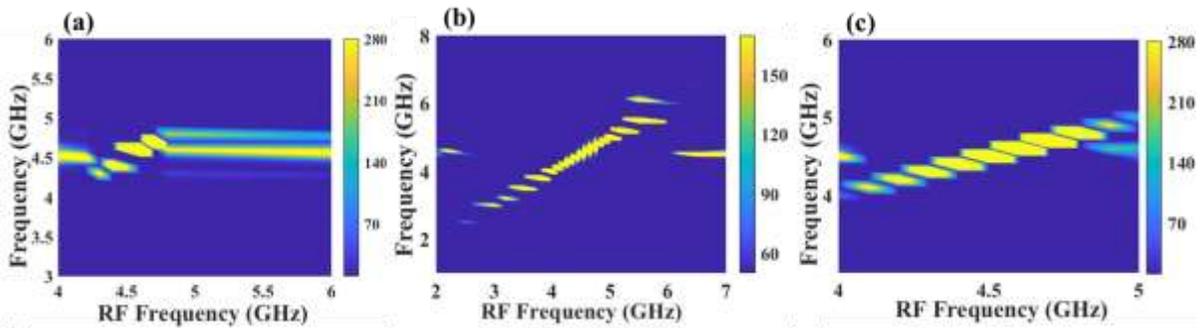

***Figure S3:*** *PSD of the time-dependent z-component of the magnetization as a function of frequency (of the magnetization signal) and the frequency of the external RF field for different RF field magnitudes and directions (with respect to single nano-constriction SHNO schematic of Fig. 2 (a) of the main text): (a) 5 Oe, along x axis, (b) 50 Oe, along x axis, (c) 50 Oe in the same direction as the DC field (within x-y plane, 24° with respect to y axis).*

**Supplementary Section S1: COMSOL Simulations**

Current density and Oersted field profiles of the single nano constriction SHNO device, obtained through COMSOL simulations, have been shown in Figure S1. Details of the structure simulated with thickness values of the layers and different parameter values are provided in Section 2 of the main text.

**Supplementary Section S2: Power Spectral Density Plots**

The plots of Fig. 2 (e)-(h) of the main text are based on power spectral density (PSD) plots of Fig. S2 and Fig. S3, as obtained through micromagnetic simulation of the single nano constriction SHNO device of Fig. 2 (a). The frequency (plotted along y axis of the PSD plots) at which PSD is maximum for a given value of DC field magnitude or DC current or frequency of RF field is plotted as a function of these quantities in Fig. 2 (e)-(h) of the main text.